\documentclass[twocolumn,conference]{IEEEtran}
\usepackage[T1]{fontenc}
\usepackage[latin9]{inputenc}
\usepackage{amsmath}
\usepackage{amssymb}
\usepackage{stackrel}
\usepackage{graphicx}
\usepackage[unicode=true,
 bookmarks=false,
 breaklinks=false,pdfborder={0 0 0},pdfborderstyle={},backref=false,colorlinks=false]
 {hyperref}
\hypersetup{pdftitle={Your Title},
 pdfauthor={Your Name},
 pdfpagelayout=OneColumn, pdfnewwindow=true, pdfstartview=XYZ, plainpages=false}

\makeatletter
\usepackage[caption=false,font=footnotesize]{subfig}

\makeatother

\begin{document}
\title{Multistatic Integrated Sensing and Communication System in Cellular
Networks}
\author{\IEEEauthorblockN{Zixiang Han, Lincong Han, Xiaozhou Zhang, Yajuan Wang, Liang Ma, Mengting
Lou, Jing Jin, Guangyi Liu}\IEEEauthorblockA{Future Research Laboratory\\
China Mobile Research Institute\\
100053, Beijing\\
Email: hanzixiang@chinamobile.com}}
\maketitle
\begin{abstract}
A novel multistatic multiple-input multiple-output (MIMO) integrated
sensing and communication (ISAC) system in cellular networks is proposed.
It can make use of widespread base stations (BSs) to perform cooperative
sensing in wide area. This system is important since the deployment
of sensing function can be achieved based on the existing mobile communication
networks at a low cost. In this system, orthogonal frequency division
multiplexing (OFDM) signals transmitted from the central BS are received
and processed by each of the neighboring BSs to estimate sensing object
parameters. A joint data processing method is then introduced to derive
the closed-form solution of objects position and velocity. Numerical
simulation shows that the proposed multistatic system can improve
the position and velocity estimation accuracy compared with monostatic
and bistatic system, demonstrating the effectiveness and promise of
implementing ISAC in the upcoming fifth generation advanced (5G-A)
and sixth generation (6G) mobile networks.
\end{abstract}


\section{Introduction}

Sensing is one of the key techniques for the upcoming fifth generation
advanced (5G-A) and sixth generation (6G) mobile networks \cite{dong2022sensing}.
Its current version focuses on the positioning of devices accessed
to the network \cite{peisa20205g}, \cite{lin2022overview}. This
limits the usage of sensing in various scenarios with massive device-free
sensing objects. To extend the application of sensing and enhance
its capability, integrated sensing and communication (ISAC) system
is recently proposed where the hardware, the spectrum and the waveform
of communication are also utilized for sensing purpose \cite{cui2021integrating},
\cite{liu2022integrated}. Therefore, considerable integration gain
can be envisioned on cost, spectral efficiency and energy efficiency
in ISAC systems. On the other hand, the sensing process needs to extract
target parameters from the propagation channel while the communication
process aims to correctly transfer the information via the channel
\cite{zhang2021overview}, \cite{hua2023optimal}. Hence, mutual benefits
between sensing and communication may be achieved by fully exploiting
results in both functions. These natural advantages make the deployment
of ISAC system more desired in the existing mobile networks \cite{zhang2020perceptive}.

To construct a multiple-input multiple-output (MIMO) ISAC system,
one simple approach is using a single base station (BS) as transceiver
to sense the surrounding environment and this is referred to as \textit{monostatic
sensing} \cite{lu2022degrees}, \cite{pucci2022system}. By transmitting
the communication signal, which is normally chosen as orthogonal frequency
division multiplexing (OFDM) signal, the BS can receive the reflected
signal from the sensing objects in the line-of-sight (LOS) scenario
\cite{barneto2019full}. Then the object parameters, including position
and radial velocity, can be obtained by estimating the angle of departure
(AoD), angle of arrival (AoA), time delay and Doppler frequency shift,
etc \cite{gu2015delay}. However, a full-duplex BS is required to
accomplish the monostatic sensing, which is a currently immature technology
\cite{wang2021sensing}. In the existing mobile networks with time-division
duplexing mode, an additional receiver can be implemented with physical
isolation to the transmitter on the same BS while the size, cost and
complexity of BSs are significantly increased. Moreover, the self-interference
signal due to BS own transmission has much larger power than the reflected
signal so that it has to be suppressed by digital cancellation technique
\cite{li2014investigation}. Therefore, it is a great burden for the
operators to deploy monostatic sensing in the mobile networks.

To address the issues in monostatic sensing, \textit{bistatic sensing}
utilizing two BSs is proposed \cite{cui2018search}, \cite{leyva2022two}
where one BS acts as transmitter with the other one being receiver
\cite{leyva2021cooperative}. To estimate the parameters of objects,
information of BS location is needed \cite{Pucci2022}. Owing to the
long distance between BSs, self-interference is automatically addressed
with no further hardware change necessary. Therefore, bistatic sensing
is more promising for the practical implementation of ISAC. However,
similar to the monostatic sensing, the bistatic sensing is unable
to fully recover the information of velocity \cite{Pucci2022}. To
overcome the challenges in both monostatic and bistatic sensing system,
in this work we propose a novel multistatic ISAC system in cellular
networks. In this system, multiple BSs are utilized to perform \textit{multistatic
sensing} where the sensing results of BS receivers can be jointly
processed. It can also take advantages of the existing mobile networks
to cover wide area and accelerate the deployment of sensing function.
To the best of our knowledge, an analysis of the multistatic sensing
in cellular networks is yet to be performed in the literature. The
specific contributions of this work are:

1) Proposing a novel multistatic MIMO ISAC system in cellular networks.
This system overcomes the challenges in monostatic and bistatic sensing
and well satisfies the requirement of large-scale ISAC implementation
for operators.

2) Proposing a low-complexity sensing method with joint data processing
to estimate the position and velocity of sensing objects. A BS scheduling
scheme is also provided.

3) Demonstrating the proposed multistatic ISAC system by simulating
the root mean square error (RMSE) of position and velocity estimation.
It is shown that the estimation accuracy in multistatic sensing can
be improved when compared with monostatic and bistatic sensing system.

\textit{Organization}: Section II formulates the multistatic ISAC
system model in cellular networks. Section III descibes an efficient
method to estimate the parameters of sensing objects. Section IV provides
numerical simulation results to demonstrate effectiveness of the multistatic
ISAC system. Section V concludes the work.

\textit{Notation}: Bold lower and upper case letters denote vectors
and matrices, respectively. Upper case letters in calligraphy denote
sets. Letters not in bold font represent scalars. $\left|a\right|$
and $E\left[a\right]$ refer to the modulus and expectation of a scalar
$a$. $\left[\mathbf{a}\right]_{i}$ and $\left\Vert \mathbf{a}\right\Vert $
refer to the $i$th element and $l_{2}-$norm of vector $\mathrm{\mathbf{a}}$,
respectively. $\mathrm{\mathbf{A}}^{T}$, $\mathrm{\mathbf{A}}^{H}$,
$\mathrm{\left[\mathbf{A}\right]}_{i,j}$ refer to the transpose,
conjugate transpose, and $\left(i,j\right)$th entry of matrix $\mathrm{\mathbf{A}}$,
respectively. $\mathbb{C}$ denotes the complex number set. $j=\sqrt{-1}$
denotes imaginary unit. 

\section{System Model}

\begin{figure}[t]
\begin{centering}
\textsf{\includegraphics[width=8cm]{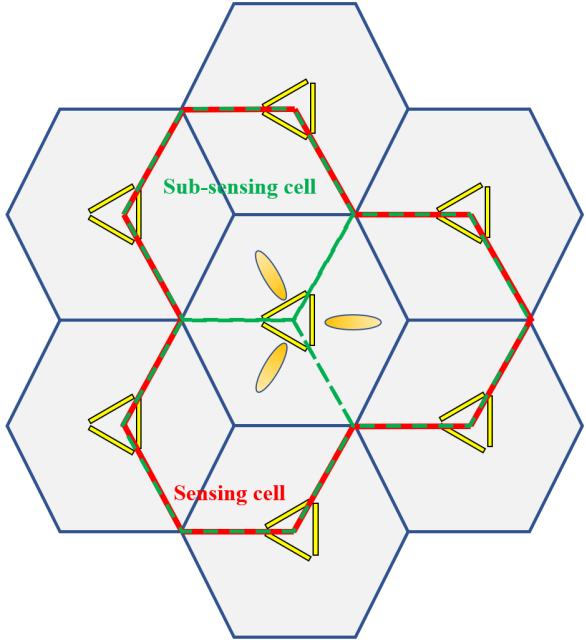}}
\par\end{centering}
\caption{Illustration of sensing cell in multistatic ISAC system based on cellular
network.\label{fig:Multistatic}}
\end{figure}

As depicted in Fig. \ref{fig:Multistatic}, we consider a fragment
of cellular network consisting of seven BSs where each BS has $3N$
antennas with every $N$ antennas cover a sector in its own cell.
To perform multistatic sensing in this fragment, the central BS (referred
to as transmitter) in Fig. \ref{fig:Multistatic} transmits signal
to its surrounding environment in the downlink mode while the reflected
signals from objects are received by $K=6$ BSs at the neighboring
cells (referred to as receiver) in the uplink mode. Therefore, six
sectors facing towards the central BS together with the central cell
form a large multistatic sensing cell (referred to as sensing cell
and enclosed by red line in Fig. \ref{fig:Multistatic}). It should
be noted that the structure of the sensing cell can be duplicated
in any cellular network, making it simple to be widely deployed.

The multistatic MIMO sensing system can be decomposed onto multiple
pairs of bistatic sensing system, then the time-domain system model
with the $k$th receiver in the sensing cell can be written as
\begin{align}
\mathbf{y}_{k}\left(t\right)= & \stackrel[l=1]{L}{\sum}\beta_{k,l}\delta\left(t-\tau_{k,l}\right)e^{j2\pi f_{D,k,l}t}\cdot\nonumber \\
 & \mathbf{a}_{R,k}\left(\Omega_{R,k,l}\right)\mathbf{a}_{T}^{T}\left(\Omega_{T,l}\right)\mathbf{x}\left(t\right)+\mathbf{n}_{k}\left(t\right),\:k=1,2,...,K,\label{eq:System Model TD}
\end{align}
where $L$ is the number of sensing objects, $\beta_{k,l}\in\mathbb{C}$
is the channel gain of the $l$th path, $\delta\left(t\right)$ is
the Dirac delta function and $\tau_{k,l}=\frac{d_{k,l}}{c}$ is the
time delay with $d_{k,l}$ and $c$ referring to the signal propagation
distance and light speed. Specifically, $d_{k,l}$ is given by $d_{k,l}=d_{T,l}+d_{R,k,l}$
where $d_{T,l}$ and $d_{R,k,l}$ are the propagation distance from
the transmitter to the $l$th object and from the $l$th object to
the $k$th receiver. $f_{D,k,l}=\frac{v_{\parallel,k,l}f_{c}}{c}$
is the Doppler frequency of the $l$th object with $v_{\parallel,k,l}$
and $f_{c}$ denoting radial velocity and carrier frequency respectively.
In particular, $\hat{v}_{\parallel,k,l}=\hat{v}_{\parallel,R,k,l}+\hat{v}_{\parallel,T,l}$
where $\hat{v}_{\parallel,R,k,l}$ and $\hat{v}_{\parallel,T,l}$
are the radial velocity of the $l$th object with respect to the $k$th
receiver and the transmitter \cite{willis2005bistatic}. $\mathbf{a}_{R,k}\left(\Omega_{R,k,l}\right)\in\mathbb{C}^{N\times1}$
and $\mathbf{a}_{T}\left(\Omega_{T,l}\right)\in\mathbb{C}^{N\times1}$
are the steering vectors of the receive antennas and the transmit
antennas respectively, $\Omega_{R,k,l}=\left(\theta_{R,k,l},\phi_{R,k,l}\right)$
and $\Omega_{T,l}=\left(\theta_{T,l},\phi_{T,l}\right)$ are angle
of arrival (AoA) and angle of departure (AoD) of the $l$th path with
$\theta$ and $\phi$ representing the elevation and azimuth angles
in spherical coordinates respectively. $\mathbf{x}\left(t\right)\in\mathbb{C}^{N\times1}$
is the transmit signal and $\mathbf{y}_{k}\left(t\right)\in\mathbb{C}^{N\times1}$
is the receive signal of the $k$th receiver. $\mathbf{n}_{k}\left(t\right)\in\mathbb{C}^{N\times1}$
is the additive noise at the $k$th receiver. 

For multistatic system with OFDM transmission scheme, we can demodulate
the receive signal in the frequency domain by performing fast Fourier
transform (FFT) on \eqref{eq:System Model TD}, then the receive OFDM
symbol on the $n_{c}$th subcarrier ($n_{c}=0,1,...,N_{c}-1$ with
$N_{c}$ being the total number of subcarriers) at the $n_{s}$th
symbol period ($n_{s}=0,1,...,N_{s}-1$ with $N_{s}$ being the total
number of OFDM symbols) is given by \cite{zhang2021overview}
\begin{align}
\tilde{\mathbf{y}}_{k}\left(f,t\right)\mid_{\begin{array}{c}
f=n_{c}f_{\Delta}\\
t=n_{s}T_{s}
\end{array}}= & \stackrel[l=1]{L}{\sum}\beta_{k,l}e^{j2\pi T_{s}f_{D,k,l}n_{s}}e^{-j2\pi\tau_{k,l}f_{\Delta}n_{c}}\cdot\nonumber \\
 & \mathbf{a}_{R,k}\left(\Omega_{R,k,l}\right)\mathbf{a}_{T}^{T}\left(\Omega_{T,l}\right)\tilde{\mathbf{x}}\left(n_{c}f_{\Delta},n_{s}T_{s}\right)\nonumber \\
 & +\tilde{\mathbf{n}}_{k}\left(n_{c}f_{\Delta},n_{s}T_{s}\right),\:k=1,2,...,K,\label{eq:System Model FD}
\end{align}
where $T_{s}$ is the OFDM symbol period, $f_{\Delta}$ is the subcarrier
spacing, $\tilde{\mathbf{x}}$ is the transmit modulated symbols on
subcarriers and $\tilde{\mathbf{n}}_{k}$ is the additive noise in
frequency domain. It should be noted that to estimate the channel
parameters related to the sensing objects, i.e. $f_{D,k,l}$, $\tau_{k,l}$,
and $\Omega_{R,k,l}$, for $k=1,2,...,K$ and $l=1,2,...,L$, the
transmit symbols $\tilde{\mathbf{x}}$ should be known at the receiver.

In this work, we consider the multistatic sensing on the azimuth plane
($\theta=90^{\circ}$) as an illustrative example to demonstrate our
proposed system, which will be detailed in the following sections.

\section{Multistatic Sensing}

In this section, we provide the solution of position and velocity
of sensing objects in multistatic system. From operators' perspective,
conventional method such as multiple signal classification (MUSIC)
\cite{schmidt1986multiple} and estimating signal via rotational invariance
techniques (ESPRIT) \cite{roy1989esprit} algorithms are not preferred
since eigenvalue decomposition in these methods cannot be efficiently
processed by BS. Therefore, we propose a low-complexity method to
estimate the channel parameters and derive the position as well as
the velocity of sensing objects.

\subsection{Low-complexity Sensing Method }

We start by estimating the AoA of receive signal in the time-domain
\eqref{eq:System Model TD}\textsf{. }Assume the AoA remains unchanged
within $N_{s}$ OFDM symbols period, we compensate \eqref{eq:System Model TD}
by using the steering vector of receive MIMO antenna array $\mathbf{a}_{R,k}\left(\phi\right)$
with angle variable $\phi$
\begin{equation}
g_{k}\left(\phi,t\right)=\mathbf{a}_{R,k}^{H}\left(\phi\right)\mathbf{y}_{k}\left(t\right),\:\phi\in\Phi{}_{k},\:t\in\left[0,N_{s}T_{s}\right]\label{eq:phase compensation}
\end{equation}
where $\Phi{}_{k}$ is the angle range for the sector of the $k$th
receiver in the sensing cell. Then we calculate the sum power of \eqref{eq:phase compensation}
at each angle during $N_{s}$ OFDM symbols period as
\begin{equation}
h_{k}\left(\phi\right)=\int_{0}^{N_{s}T_{s}}\left|g_{k}\left(\phi,t\right)\right|\mathrm{d}t.\label{eq:angle sum power}
\end{equation}
Generally when $\phi=\phi_{R,k,l}$, $l=1,2,...,L$, the steering
vector of receive antenna $\mathbf{a}_{R,k}\left(\phi\right)$ in
\eqref{eq:System Model TD} can be perfectly compensated so that \eqref{eq:angle sum power}
reaches local maximum value. Therefore, AoA $\hat{\phi}_{R,k}$ can
be estimated by finding the peaks of $h_{k}\left(\phi\right)$ as
\begin{equation}
\hat{\Phi}_{R,k}=\left\{ \hat{\phi}_{R,k}\mid h_{k}>\epsilon_{\phi},\frac{\mathrm{d}h_{k}}{\mathrm{d}\phi}=0,\frac{\mathrm{d^{2}}h_{k}}{\mathrm{d^{2}\phi}}<0\right\} \label{eq:AoA peak}
\end{equation}
where $\epsilon_{\phi}$ is the threshold value which is related to
the noise and interference power.

Next we estimate the time delay and radial velocity based on the receive
OFDM symbols \eqref{eq:System Model FD} and the estimated AoA. We
firstly remove the information of transmit symbols $\tilde{\mathbf{x}}$
from the receive symbol $\tilde{\mathbf{y}}_{k}$. Specifically, the
receive symbols $\tilde{\mathbf{y}}_{k}$ divided by the $i$th entry
in $\tilde{\mathbf{x}}$ is written as
\begin{align}
\tilde{\mathbf{h}}_{k,i}\left(n_{s},n_{c}\right)=\frac{\tilde{\mathbf{y}}_{k}}{\left[\tilde{\mathbf{x}}\right]_{i}}= & \stackrel[l=1]{L}{\sum}\beta_{k,l}e^{j2\pi T_{s}f_{D,k,l}n_{s}}e^{-j2\pi\tau_{k,l}f_{\Delta}n_{c}}\cdot\nonumber \\
 & \left[\mathbf{a}_{T}\left(\phi_{T,l}\right)\right]_{i}\mathbf{a}_{R,k}\left(\phi_{R,k,l}\right)+\bar{\mathbf{n}}_{k,i},\nonumber \\
 & k=1,2,...,K,\:i=1,2,...,N,\label{eq:entry-wise division}
\end{align}
where $\left[\mathbf{a}_{T}\left(\phi_{T,l}\right)\right]_{i}$ in
$\tilde{\mathbf{h}}_{k,i}\in\mathbb{C}^{N\times1}$ can be combined
into channel gain $\beta_{k,l}$. Then by compensating \eqref{eq:entry-wise division}
with the steering vector of receive antenna for each angle in $\hat{\Phi}_{R,k}$,
we have
\begin{align}
\tilde{g}_{k,i}\left(n_{s},n_{c}\right)= & \mathbf{a}_{R,k}^{H}\left(\phi\right)\tilde{\mathbf{h}}_{k,i}\left(n_{s},n_{c}\right),\:\phi\in\hat{\Phi}_{R,k},\label{eq:FD phase compensation}
\end{align}
where $\tilde{g}_{k,i}$ is a function of subcarrier index and symbol
index on which we can perform two dimension (2D) discrete Fourier
transform (DFT) \cite{strum2011waveform}
\begin{align}
\mathbf{G}_{k,i}\left(n_{s},n_{c}\right)= & \mathbf{F}\tilde{g}_{k,i}\left(n_{s},n_{c}\right),\:i=1,2,...,N,\label{eq:2DDFT}
\end{align}
where $\mathbf{F}\in\mathbb{C}^{N_{s}\times N_{c}}$ is the 2D-DFT
matrix with $\left[\mathbf{F}\right]_{p,q}=\frac{1}{N_{c}N_{s}}e^{-j\left(p-\frac{N_{s}}{2}\right)\frac{2\pi}{N_{s}}n_{s}}e^{j\left(q-1\right)\frac{2\pi}{N_{c}}n_{c}}$
($N_{s}$ is assumed to be even). By taking \eqref{eq:FD phase compensation}
into \eqref{eq:2DDFT}, the phase brought by time delay and radial
velocity can be offset when
\begin{equation}
T_{s}f_{D}=\frac{\left(\hat{p}_{k,i}-\frac{N_{s}}{2}\right)}{N_{s}}.\label{eq:Doppler Offset}
\end{equation}
\begin{equation}
\tau f_{\Delta}=\frac{\left(\hat{q}_{k,i}-1\right)}{N_{c}},\label{eq:Time Delay Offset}
\end{equation}
where $\left(\hat{p}_{k,i},\hat{q}_{k,i}\right)$ is the index of
peak in $\mathbf{G}_{k,i}$. It should be noted that the steps from
\eqref{eq:entry-wise division} to \eqref{eq:Time Delay Offset} are
the same for each transmit OFDM stream $i=1,2,...,N$ so that we define
average peak index $\left(\hat{p}_{k},\hat{q}_{k}\right)=\frac{1}{N}\stackrel[i=1]{N}{\sum}\left(\hat{p}_{k,i},\hat{q}_{k,i}\right)$
as the estimated peak index. The propagation distance $d$ and radial
velocity $v_{\parallel}$ of the objects can then be solved as
\begin{equation}
\hat{v}_{\parallel,k}=\frac{\left(\hat{p}_{k}-\frac{N_{s}}{2}\right)c}{N_{s}T_{s}f_{c}},\label{eq:radial velocity}
\end{equation}
\begin{equation}
\hat{d}_{k}=\frac{\left(\hat{q}_{k}-1\right)c}{N_{c}f_{\Delta}}.\label{eq:distance}
\end{equation}
By finding all peaks in $\mathbf{G}_{k,i}$ for each of the given
angle in $\hat{\Phi}_{R,k}$, the estimated channel parameters of
all sensing objects $\left(\hat{\phi}_{R,k},\hat{d}_{k},\hat{v}_{\parallel,k}\right)$
can be obtained. 

We also provide here brief expressions of the estimation error on
$\left(\hat{\phi}_{R,k},\hat{d}_{k},\hat{v}_{\parallel,k}\right)$.
For the estimation of AoA, the error $\phi^{e}=\left|\hat{\phi}_{R,k}-\phi_{R,k}\right|$
depends on step size of $\phi$, denoted by $\Delta\phi$, and can
be given by $E\left[\phi^{e}\right]=\frac{\Delta\phi}{4}$. This is
because $\phi^{e}$ is uniformly distributed within $\left[0,\frac{\Delta\phi}{2}\right]$.
Similarly, the resolution of radial velocity and propagation distance
are respectively given by $\Delta v_{\parallel}=\frac{c}{N_{s}T_{s}f_{c}}$
and $\Delta d=\frac{c}{N_{c}f_{\Delta}}$. Therefore, the estimation
error are written as $E\left[v_{\parallel}^{e}\right]=\frac{c}{4N_{s}T_{s}f_{c}}$
and $E\left[d^{e}\right]=\frac{c}{4N_{c}f_{\Delta}}$ which are inversely
proportional to the total OFDM symbol period $N_{s}T_{s}$ and system
bandwidth $N_{c}f_{\Delta}$ respectively.

\subsection{Position and Velocity Estimation}

In this subsection, we derive the closed-form solution of position
and velocity for sensing objects based on the estimation of AoA $\hat{\phi}_{R,k}$,
propagation distance $\hat{d}_{k}$ and radial velocity $\hat{v}_{\parallel,k}$
of multiple receivers.
\begin{figure}[t]
\begin{centering}
\textsf{\includegraphics[width=8cm]{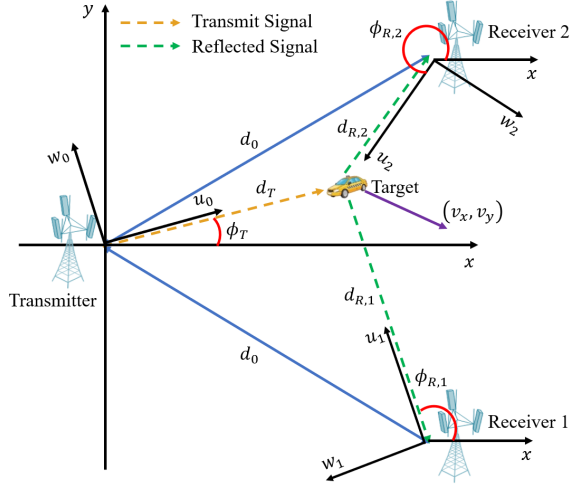}}
\par\end{centering}
\caption{Illustration of position and velocity estimation.\label{fig:Position and Velocity}}
\end{figure}

Assume the coordinate of transmitter is $\left(x_{0},y_{0}\right)$
and the distance between the transmitter and receiver is $d_{0}$,
then the location of receivers are given by $\left(x_{k},y_{k}\right)=\left(x_{0}+d_{0}\mathrm{cos}\left(\frac{2k\pi-1}{K}\right),y_{0}+d_{0}\mathrm{sin}\left(\frac{2k\pi-1}{K}\right)\right)$
for $k=1,2,...,K$. Utilizing the Law of Cosines together with AoA
$\hat{\phi}_{R,k}$ and propagation distance $\hat{d}_{k}$, the distance
between the object and $k$th receiver can be derived as
\begin{equation}
\hat{d}_{R,k}=\frac{\hat{d}_{k}^{2}-d_{0}^{2}}{2\left(\hat{d}_{k}+d_{0}\mathrm{cos}\left(\frac{2k\pi-1}{K}-\hat{\phi}_{R,k}\right)\right)}.\label{eq:dr}
\end{equation}
Therefore, the position of the object estimated by the $k$th receiver
is given by
\begin{equation}
\hat{x}_{k}=x_{k}+\hat{d}_{R,k}\mathrm{cos}\left(\hat{\phi}_{R,k}\right),\label{eq:x estimate}
\end{equation}
\begin{equation}
\hat{y}_{k}=y_{k}+\hat{d}_{R,k}\mathrm{sin}\left(\hat{\phi}_{R,k}\right),\label{eq:y estimate}
\end{equation}
which are sent to server for joint data processing. In the single
object case, the mean position can be obtained by averaging the position
estimation results from multiple receivers as
\begin{equation}
\left(\hat{x},\hat{y}\right)=\frac{1}{K}\stackrel[k=1]{K}{\sum}\left(\hat{x}_{k},\hat{y}_{k}\right).\label{eq:Mean Position}
\end{equation}
 Meanwhile, the estimated AoD is given by
\begin{equation}
\hat{\phi}_{T}=\mathrm{arctan}\left(\frac{\hat{y}-y_{0}}{\hat{x}-x_{0}}\right).\label{eq:AoD}
\end{equation}
It should be noted that deriving the position of object by using two
estimated AoA from two receivers cannot be applied in the scenario
with multiple objects since the server is unable to properly match
multiple AoAs from different receivers without additional information.
In this scenario, clustering the position estimation results from
different receivers based on the minimum Euclidean distance should
be firstly performed. The mean position of sensing objects are then
obtained by \eqref{eq:Mean Position} for each cluster. 

Next we derive the velocity of sensing objects based on the estimation
results of AoA $\hat{\phi}_{R,k}$, AoD $\hat{\phi}_{T}$ and radial
velocity $\hat{v}_{\parallel,k}$. As shown in Fig. \ref{fig:Position and Velocity},
we set up local coordinate $\left(u,w\right)$ with the origin being
the position of transmitter or receiver and the positive $u$-axis
pointing from the origin to the object. The coordinate transformation
from global coordinate $\left(x,y\right)$ to local coordinate $\left(u,w\right)$
can be expressed by matrix
\begin{equation}
\mathbf{T}\left(\hat{\phi}\right)=\left[\begin{array}{cc}
\mathrm{cos}\left(\hat{\phi}\right) & \mathrm{sin}\left(\hat{\phi}\right)\\
-\mathrm{sin}\left(\hat{\phi}\right) & \mathrm{cos}\left(\hat{\phi}\right)
\end{array}\right].\label{eq:coordinate transform}
\end{equation}
Then the velocity in global coordinate and in local coordinate for
the $k$th receiver and the transmitter are respectively related by
\begin{equation}
\left[\begin{array}{c}
\hat{v}_{\parallel,R,k}\\
\hat{v}_{\perp,R,k}
\end{array}\right]=\mathbf{T}\left(\hat{\phi}_{R,k}\right)\left[\begin{array}{c}
\hat{v}_{x}\\
\hat{v}_{y}
\end{array}\right],\label{eq:velocity calculation receiver}
\end{equation}
\begin{equation}
\left[\begin{array}{c}
\hat{v}_{\parallel,T}\\
\hat{v}_{\perp,T}
\end{array}\right]=\mathbf{T}\left(\hat{\phi}_{T}\right)\left[\begin{array}{c}
\hat{v}_{x}\\
\hat{v}_{y}
\end{array}\right],\label{eq:velocity calculation transmitter}
\end{equation}
where $\hat{v}_{x}$ and $\hat{v}_{y}$ are estimated velocity of
object in the $x$-axis and $y$-axis direction in global coordinate,
$\hat{v}_{\perp,R,k}$ and $\hat{v}_{\perp,T}$ are the normal velocity
with respect to the $k$th receiver and the transmitter. By taking
$\hat{v}_{\parallel,R,k}$ and $\hat{v}_{\parallel,T}$ in \eqref{eq:velocity calculation receiver}
and \eqref{eq:velocity calculation transmitter} into \eqref{eq:radial velocity}
we have
\begin{align}
\hat{v}_{\parallel,k}= & \left(\mathrm{cos}\left(\hat{\phi}_{R,k}\right)+\mathrm{cos}\left(\hat{\phi}_{T}\right)\right)\hat{v}_{x}\nonumber \\
 & +\left(\mathrm{sin}\left(\hat{\phi}_{R,k}\right)+\mathrm{sin}\left(\hat{\phi}_{T}\right)\right)\hat{v}_{y}\label{eq:vel with vx vy}
\end{align}
To solve $\hat{v}_{x}$ and $\hat{v}_{y}$, at least two radial velocity
results, denoted as $\hat{v}_{\parallel,k_{1}}$ and $\hat{v}_{\parallel,k_{2}}$,
estimated from the $k_{1}$th and $k_{2}$th ($k_{1}\neq k_{2}$)
receivers are required to perform joint processing. Alternatively,
$\mathrm{sin}\left(\hat{\phi}_{T}\right)$ and $\mathrm{cos}\left(\hat{\phi}_{T}\right)$
in \eqref{eq:vel with vx vy} can be canceled out by using three radial
velocity results for the derivation of $\hat{v}_{x}$ and $\hat{v}_{y}$,
which can avoid the usage of AoD estimation results. Therefore, the
velocity of sensing objects can be fully recovered in the multistatic
system.

\subsection{BS Scheduling Scheme}

\begin{figure*}[t]
\begin{centering}
\textsf{\includegraphics[width=14cm]{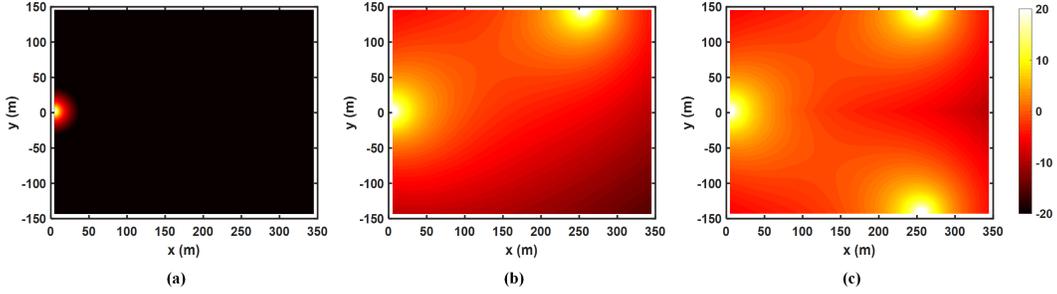}}
\par\end{centering}
\caption{Simulated results relating the receive SNR to the object position
in (a) monostatic, (b) bistatic and (c) multistatic sensing.\label{fig:SNR Comparison}}
\end{figure*}

It can be noticed from the above analysis that multistatic sensing
with two receivers provide enough information for obtaining the position
as well as the velocity of sensing objects. In addition, the path
gain is inversely proportional to the propagation distance square,
which can be modeled as
\begin{equation}
\mathrm{PG}=\frac{\eta G_{T}G_{R}c^{2}}{\left(4\pi\right)^{3}d_{T}^{2}d_{R}^{2}f_{c}^{2}}\label{eq:Path Loss}
\end{equation}
where $\eta$ is the radar cross-section (RCS) of sensing objects,
$G_{T}$ and $G_{R}$ are the antenna gain of transmitter and receiver.
Therefore, only the neighboring BSs with smallest $d_{R}$ can detect
the reflected signal with significant power from the sensing objects.
To avoid wasting BS resources, we discretize the sensing cell into
three sub-sensing cells, each of which can be covered by two adjacent
receivers as shown in Fig. \ref{fig:Multistatic} (enclosed by green
dashed line). When sensing function is required in an area, two receivers
in the corresponding sub-sensing cell are scheduled for sensing purpose.
In this way, the maximum propagation distance in the sub-sensing cell
is $2d_{0}$ when $d_{T}=d_{R}=d_{0}$ so that the $\mathrm{PG}$
can be always greater than $\frac{\eta G_{T}G_{R}c^{2}}{\left(4\pi\right)^{3}d_{0}^{4}f_{c}^{2}}$
which is associated with the minimum receive signal power.

\section{Numerical Simulation}

In the numerical simulation, multistatic sensing system in a sub-sensing
cell as shown in Fig. \ref{fig:Multistatic} is considered. To evaluate
its performance in position and velocity estimation, we assume a single
sensing object locating in the sub-sensing cell. The range of object
velocity is assumed to be $\left|v\right|\in\left[0,30\right]$ with
arbitrary direction. We consider the carrier frequency of $f_{c}=2.6$
GHz where subcarrier spacing is $f_{\Delta}=30$ kHz and the OFDM
symbol period is $T_{s}=\frac{0.5}{14}\:\mathrm{ms}=0.0357$ ms. Besides,
the number of transmit and receive antennas are $N=16$ with the total
antenna gain being $G_{T}=G_{R}=12$ dB. The distance between the
transmitter and receiver is assumed as $d_{0}=300$ m and the location
of transmitter and two receivers in the sub-sensing cell are given
by $\left(0,0\right)$ m, $\left(260,150\right)$ m and $\left(260,-150\right)$
m.

We define the signal-to-interference-plus-noise ratio (SINR) as
\begin{equation}
\mathrm{SNR}=\frac{\mathrm{PG}\cdot P_{T}}{N_{0}+I_{0}}\label{eq:SNR}
\end{equation}
where $P_{T}$ is the transmit signal power, $N_{0}$ is the noise
power given by $N_{0}=-174+10\mathrm{log_{10}}\mathrm{BW}+NF$ dBm
with $NF$ denoting noise figure, $I_{0}$ is the power of self-interference
signal defined as $I_{0}=\alpha P_{T}$ with $\alpha$ representing
the coefficient of self-interference cancellation. In the simulation,
we use $\alpha=-70$ dB for monostatic sensing and $\alpha=0$ for
bistatic and multistatic sensing. The other parameters are as follows:
$P_{T}=30$ dBm, $\eta=1$, and $NF=10$ dB.

From \eqref{eq:Path Loss} and \eqref{eq:SNR}, we can notice that
SNR is a function of both propagation distance and interference power.
Considering that SNR is the key factor of the accuracy in channel
parameter estimation and the resulting position as well as velocity
estimation in Section III, we firstly make comparison on receive SNR
between the proposed multistatic sensing and the monostatic \& bistatic
sensing. It should be noted again that in monostatic sensing, the
transmitter also receives the reflected signal of sensing object.
In Fig. $\ref{fig:SNR Comparison}$, we plot the receive SNR when
sensing object locates at various positions in the sub-sensing cell.
It can be observed that in multistatic sensing system, the maximum
SNR at the receiver maintains above $-10$ dB for arbitrary position
of sensing objects. This is because two BSs can receive the reflected
signal simultaneously and SNR is always high for at least one of them.
We also notice that SNR in monostatic sensing degrades sharply due
to the interference power and the increase of propagation distance
while SNR in bistatic sensing also decreases obviously as the object
moves away from the BSs. Therefore, our proposed multistatic sensing
system can cover larger area based on the cellular networks when compared
with monostatic and bistatic sensing.

\begin{figure}[t]
\begin{centering}
\textsf{\includegraphics[width=8cm]{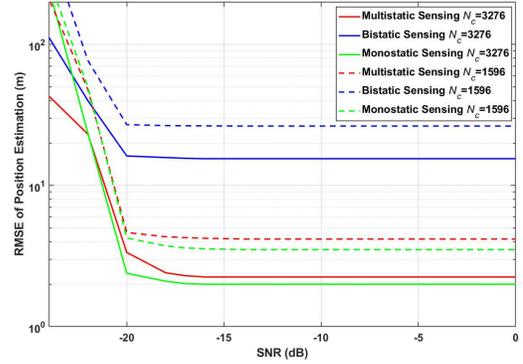}}
\par\end{centering}
\begin{centering}
(a)
\par\end{centering}
\begin{centering}
\textsf{\includegraphics[width=8cm]{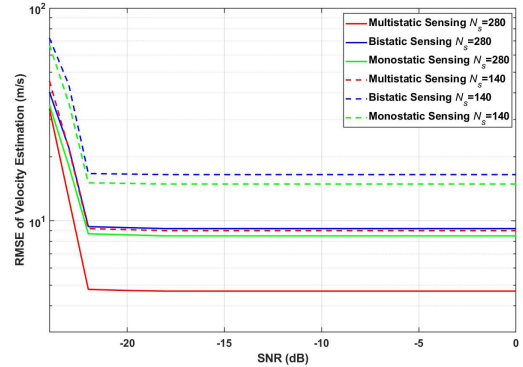}}
\par\end{centering}
\begin{centering}
(b)
\par\end{centering}
\caption{Simulated RMSE of (a) position and (b) velocity estimation in the
proposed multistatic sensing system.\label{fig:RMSE}}
\end{figure}

To further highlight the advantages of the proposed multistatic sensing
system, we simulate the root mean square error (RMSE) of estimated
position and velocity with respect to the actual values of sensing
objects as shown in Fig. \ref{fig:RMSE}, which is also benchmarked
with the performance of monostatic and bistatic sensing. For the position
estimation in Fig. \ref{fig:RMSE}(a), we can make following observations:
\textit{Firstly}, it can be observed that the RMSE of positioning
in multistatic sensing is smaller than that in bistatic sensing, which
demonstrates the effectiveness of utilizing multiple receivers to
improve the positioning accuracy. \textit{Secondly}, the RMSE of positioning
in multistatic sensing is slightly higher than that in monostatic
sensing. This is because the estimation error on distance is reduced
by half for the round-trip propagation in monostatic sensing, which
is beneficial for the position estimation. However, combining with
the results in Fig. \ref{fig:SNR Comparison}, the multistatic sensing
outperforms the monostatic sensing due to the high receive SNR for
arbitrary object position. \textit{Thirdly}, by comparing the results
with different number of subcarriers $N_{c}$, we can notice that
a larger system bandwidth can reduce the RMSE of position estimation.

We can also make several observations about the velocity estimation
as shown in Fig. \ref{fig:RMSE}(b)\textit{. }It can be noticed that
the RMSE of velocity estimation in multistatic sensing is much smaller
than that in bistatic and monostatic sensing since bistatic and monostatic
sensing can only provide the radial velocity of objects with significant
error in terms of the normal velocity. Furthermore, The velocity estimation
error in the multistatic sensing is related to the number of OFDM
symbols $N_{s}$ where longer OFDM symbol period are desired for improving
the velocity estimation accuracy.

To sum up, by using multistatic sensing in cellular network, the sensing
area can be greatly expanded and both the position and velocity estimation
accuracy can be significantly improved when compared with monostatic
and bistatic sensing.

\section{Conclusions}

In this paper, we propose a novel multistatic MIMO ISAC system in
cellular networks. The proposed system well satisfies the ISAC implementation
requirement of operators by making use of widely deployed BSs. Specifically
in this system, the single transmit BS cooperates with multiple neighboring
receive BSs whose sensing results are jointly processed to estimate
the position and velocity of sensing objects. A BS scheduling scheme
is also proposed to avoid wasting BS resources. It is shown by the
numerical simulation that the proposed system can cover larger area
and significantly reduce the position and velocity estimation error
when compared with monostatic and bistatic sensing system, demonstrating
the effectiveness of the proposed multistatic sensing and the promise
of implementing such system in the upcoming 6G mobile networks.

This proposed system can serves as an initial guidance in ISAC deployment
for operators. For future works, the design and the impact of beamforming
at both transmitter and receiver side in multistatic sensing can be
investigated. Beamforming at the transmitter can improve SNR by steering
the main lobe to point to the sensing objects while the beamforming
matrix is unknown to the receiver without additional notification.
Therefore, the impact of unknown beamforming matrix on data decoding
and channel parameters estimation requires further investigation.

\bibliographystyle{IEEEtran}

\end{document}